\documentclass[runningheads]{llncs}
\usepackage{graphicx}
\usepackage{amsmath,amssymb} 
\usepackage{color}
\usepackage[width=122mm,left=12mm,paperwidth=146mm,height=193mm,top=12mm,paperheight=217mm]{geometry}
\usepackage{graphicx}
\usepackage{cite}
\usepackage{subcaption}
\usepackage{amsmath}
\DeclareMathOperator*{\argmax}{argmax} 
\usepackage{algorithm}
\usepackage[noend]{algpseudocode}
\usepackage[colorlinks = true,
            linkcolor = blue,
            urlcolor  = blue,
            citecolor = blue,
            anchorcolor = blue]{hyperref}

\begin{document}
\pagestyle{headings}
\mainmatter
\def\ECCV16SubNumber{***}  

\title{Predicting Eye Fixations Under Distortion Using Bayesian Observers} 

\titlerunning{PSY380E VISION SYSTEMS}

\authorrunning{Zhengzhong Tu}

\author{Zhengzhong Tu}
\institute{University of Texas at Austin}

\maketitle

\begin{abstract}
Visual attention is very an essential factor that affects how human perceives visual signals. This report investigates how distortions in an image could distract human's visual attention using Bayesian visual search models, specifically, Maximum-a-posteriori (MAP) \cite{findlay1982global}\cite{eckstein2001quantifying} and Entropy Limit Minimization (ELM) \cite{najemnik2009simple}, which predict eye fixation movements based on a Bayesian probabilistic framework. Experiments on modified MAP and ELM models on JPEG-compressed images containing blocking or ringing artifacts were conducted and we observed that compression artifacts can affect visual attention. We hope this work sheds light on the interactions between visual attention and perceptual quality. 


\end{abstract}

\section{Introduction}

Visual attention (VA) plays a critical role in human vision systems (HVS) given the human brain's limited capacity of selected neural processing as well as the narrow and high resolution foveated receptive field in the retina. Visual attention is, by definition, the behavioral and cognitive process of selectively concentrating on a discrete aspect of information, whether deemed subjective or objective, while ignoring other perceivable information \cite{wiki:xxx}. While visual attention guides the anatomical structures in HVS to capture more detailed information for important scenes in the context of some specific task, the main interest is on the mechanism underlying this guidance.

Recent decades have witnessed a growing interest in exploring the mechanisms of visual attention. Studies on modeling visual attention, however, is among many facets of scientific research such as psychology, neuroscience, biology, computer vision, and robotics. Generally, there are two primary roots of motivation in the taxonomy of visual attention \cite{Tsotsos:2011}: one for the uses of attentive methods in computer vision and another for the development of attention models in the biological vision community. However, although both have developed independently, and indeed, the use of attention appears in the computer vision literature before most biological models, the major point of intersection is the class of computational models. A computational model of visual attention, according to \cite{itti2001computational}, not only includes a formal or mathematical description for how attention is computed, but also has to be in a testable manner, i.e., can be tested by providing input stimuli such as image/video signals to subjects and see how the model performs by comparison. We refer the reader to \cite{frintrop2010computational} for a comprehensive review of computational models of visual attention.

Here we scope down to models only under saliency map assumptions, originated from Treisman and Gelade's \cite{treisman1980feature} ``Feature Integration Theory''. Koch and Ullman \cite{koch1987shifts} then proposed a feed-forward model to combine these features and introduced the concept of \textbf{saliency map}, which is a topographic map that represents conspicuousness of scene locations. The first complete implementation and verification of the Koch and Ullman model was proposed by Itti et al. \cite{itti1998model} and applied to both synthetic as well as natural scenes. Since then, there have been increasing interests on saliency detection or salient object detection tasks on computer vision field. 

\subsection{Saliency-based Models}
\label{sec:sbm}

In computer vision community, stimulus-driven, saliency-based attention has been a trendy research area over the past 25 years. Given the difficulty of accurately measuring or even quantifying the internal states of organisms, bottom-up attention, independent of these biological aspects, is much easier to understand and evaluate. More specifically, here we are studying visual attention models that can compute saliency maps, which characterize some parts of a scene that stand out relative to their neighboring parts. Here, we give a clear definition and description of what kind of problems this paper is approaching from a computational standpoint. These definitions generally refer to \cite{borji2013state}.

Suppose $K$ subjects have viewed a dataset consisting of $N$ images $\mathcal{I}=\{\boldsymbol{I}_i\}_{i=1}^N$. Let $L_i^k=\{\boldsymbol{p}_{ij}^k,\boldsymbol{t}_{ij}^k\}_{j=1}^{n_i^k}$ be the vector of eye fixations $\boldsymbol{p}_{ij}^k=(\boldsymbol{x}_{ij}^k,\boldsymbol{y}_{ij}^k)$ and their corresponding occurrence time $\boldsymbol{t}_{ij}^k$ for the $k$th subject over image $\boldsymbol{I}_i$. Let the number of fixations of this subject over $i$th image be $n_i^k$. The goal of saliency-based attention modeling is to find a function (stimuli-saliency mapping) $f\in \mathcal{F}$ which minimizes the error on eye fixation prediction, i.e., minimizes $\sum_{k=1}^K\sum_{i=1}^N m(f(\mathbf{I}_i^k, \mathbf{L}_i^k))$, where $m\in \mathcal{M}$ is a distance measure. An important point here is that the above definition better suits bottom-up models of overt visual attention, and may not necessarily cover some other aspects of visual attention (e.g., covert attention or top-down factors).

The problem statement described above is for general saliency research given some computer vision task or even not given any instructions to the subjects. Actually, many technologies and applications of these models have been developed for computer vision and robotics motivations such as image segmentation, image object detection, and human-robot interaction, but few have done for visual quality research purposes. Hence, we would prefer to bound our interest to image quality assessment since less attention has been paid on the variation of saliency map given the degradation of image quality. In the next section, we will articulate the models researched in this paper, as well as their applications to image quality research.

\subsection{Visual Search Model}

Visual search models have been used as very intuitive schemes to generate saliency maps in that they can directly predict the eye fixation positions, from which it is easy to construct saliency map by merely applying Gaussian-blurred linear operations. In the visual search literature, many researchers have proposed to model fixation selection strategies of searchers from Random Search \cite{engel1977visual}, Tiling Search \cite{geisler1995separation} to Feature-based Search \cite{treisman1980feature}\cite{wolfe1998can}\cite{eckstein2001quantifying}. Despite decades of research, however, relatively little is known about how humans actually select fixation locations in visual search. 

In recent years, some research works \cite{najemnik2005optimal}\cite{najemnik2008eye}\cite{najemnik2009simple}\cite{abrams2015visual} have shown that a Bayesian ideal searcher under a general probabilistic framework can perform statistically as good as human eye movements. Therefore, we would like to develop an eye fixation prediction algorithm based on the Bayesian inference model proposed by Najemnik and Geisler \cite{najemnik2005optimal}. Conceptually, this Bayesian model depicted in flowchart Fig. \ref{fig:2} works as follows: a searcher starts with some initial priors where the target would be. Then the searcher encodes the image with a foveated visual system to optimally update the priors, from which the posterior probabilities are generated. Afterward, if the posterior at some location gets large enough, the search is stopped, and the position with the largest posterior is chosen as the next fixation; if the posteriors are all under the given threshold, then the searcher uses some cost functions to optimally pick the following fixate location. Under this framework, some popular searchers like MAP \cite{findlay1982global}, ELM \cite{najemnik2009simple}, and nELM \cite{abrams2015visual} have been developed to model human eye movement strategy. However, this optimal fixation search framework was generally used for target searching tasks under natural image background, whereas not directly applicable for free-looking searching on distorted images. In the next section, we will slightly modify and redefine some denotations for implementations of the framework to better fit the context of searching free-looking eye fixations on pristine or distorted images.

\section{Eye Fixation Searchers}

\begin{figure}[t]
    \centering
    \includegraphics[width=0.7\columnwidth]{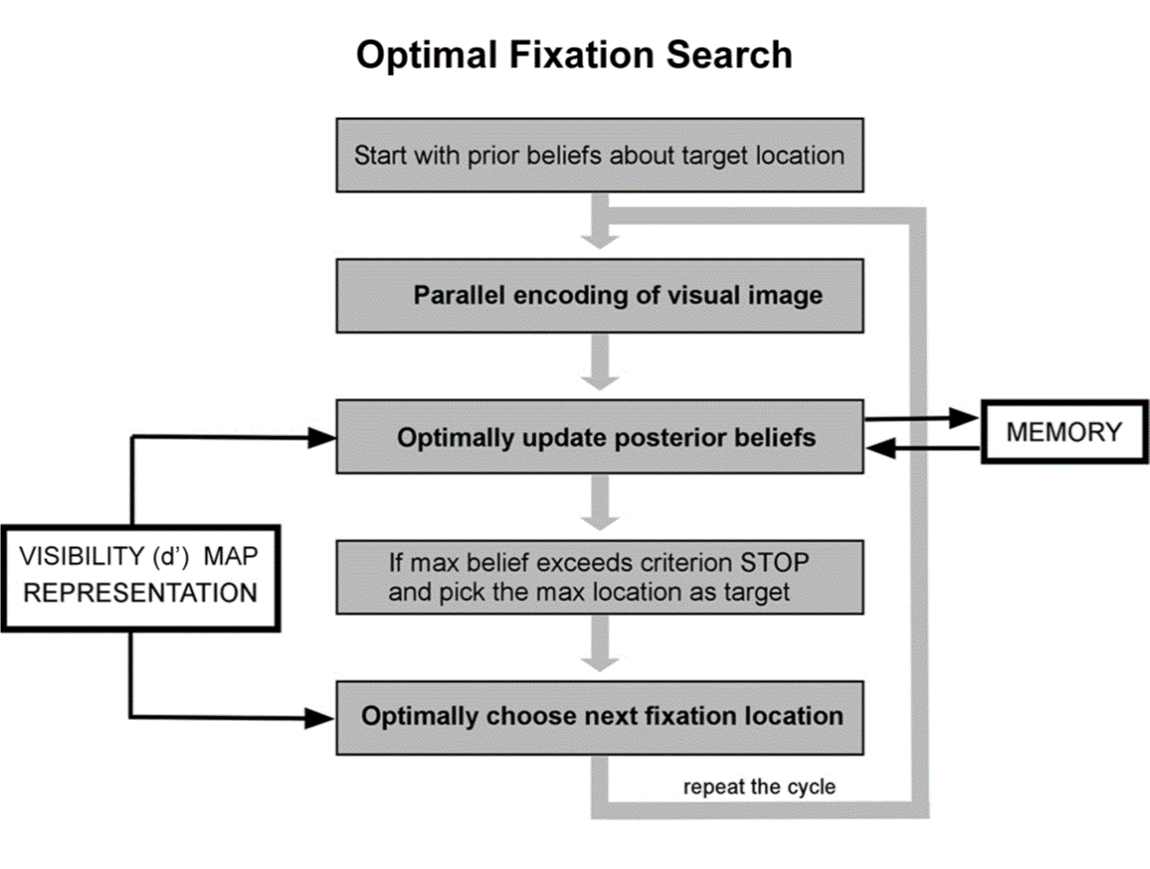}
    \caption{Flow chart for ideal Bayesian searcher in a probabilistic framework \cite{najemnik2005optimal}}
    \label{fig:2}
\end{figure}

This section describes in detail how we implemented the eye fixation search algorithm based on the Bayesian ideal search model Fig. \ref{fig:2} \cite{najemnik2005optimal}. We first define the visibility map (section \ref{sec:sbm}), image patch response (section \ref{ssec:patch}), Bayesian updating scheme (section \ref{ssec:bayes}), and the optimal selection strategy (section \ref{ssec:select}) as the most essential modules within this Bayesian framework, and then integrate them all to present the overall procedures of each simulated search trial among three representative searchers MAP \cite{findlay1982global}, ELM \cite{najemnik2009simple}, and nELM \cite{abrams2015visual} respectively in section \ref{ssec:sim}.

\subsection{Visibility Map Representation}
\label{ssec:vis}

The detrimental effect of retinal eccentricity on the visibility of a target or region was implemented by modeling detectability, $\boldsymbol{d'}$, as a function of eccentricity $\boldsymbol{\epsilon(i,k(t))}$, 

\begin{equation}
\label{eq1}
    d_{ik(t)}'=\max \Big\{0.01, \mu \exp{\Big[-\frac{\epsilon(i,k(t))^2}{2\sigma^2}\Big]}\Big\}
\end{equation}

\begin{figure}[t]
    \centering
    \begin{subfigure}[b]{0.45\textwidth}
        \includegraphics[width=\textwidth]{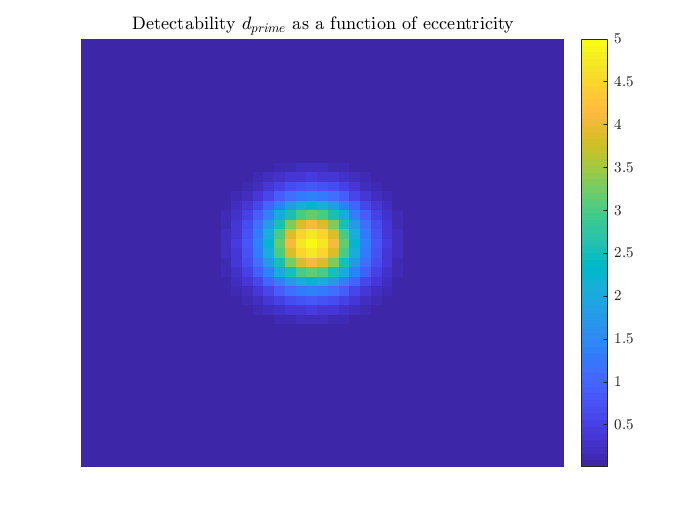}
        \caption{Visibility map at center}
        \label{fig:3(a)}
    \end{subfigure} 
    ~ 
    \begin{subfigure}[b]{0.45\textwidth}
        \includegraphics[width=\textwidth]{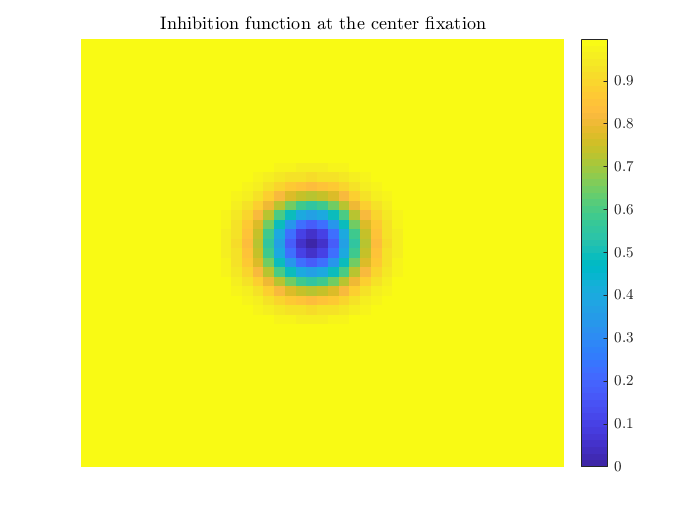}
        \caption{Inhibition function at center}
        \label{fig:3(b)}
    \end{subfigure} 
    \caption{Representation of the visibility map and inhibition function}
    \label{fig:3}
\end{figure}

Where the magnitude $\mu$ and standard deviation $\sigma$ are fit to measure the visibility map for each human subject, and $\epsilon(i,k(t))$ is simply the distance between target response location $i$ and fixation location $k(t)$. Note that we restrict the visibility value to be no less than $0.01$ for stabilization. Here for simplicity, we set the parameters to be fixed in our simulations as $\mu=5,\ \sigma=50,\ \epsilon=\|\cdot\|_2$. The simulated visibility map when fixation is at the center of the image is shown in Fig. \ref{fig:3(a)}.

\subsection{Image Patch Response}
\label{ssec:patch}

First we divie the image $I$ (resolution $W\times H$) into small patches with size $16\times 16$ pixels, and therefore we can get overall $M=(W\times H/256)$ number of patches $\{P_i\}_{i=1}^M$. We limit the potential fixation positions to only be at the center of patches to simplify both the analytic and computational load. Therefore, we get the fixations set as $k(\mathcal{T})=\{k(t)\}_{t=1}^T$, where $k(t)$ represents the location of fixation $t$. Note that here we have $M=T$.

We are now able to define the natural image response at a given image patch $P_i$. Unlike what they did for template response settings in \cite{najemnik2008eye}, here instead we choose three features highly related to region conspicuousness in order to roughly indicate the response variable $W_{ik(t)}$ at patch $P_i$ modeled as random variables sampled from Gaussian distributions:

\begin{equation}
\begin{gathered}
\label{eq2}
\mathcal{W}_{ik(t)}=\mathcal{E}(C_i,L_i,H_i)+\mathcal{N}_{ik(t)},\ \forall i=1,...,M \\
\mathcal{E}(C_i,L_i,H_i)=\alpha\cdot C_i^\beta\cdot L_i^\gamma\cdot H_i^\eta \\
\mathcal{N}_{ik(t)}\sim \mathcal{N}(0,\ 1/d_{ik(t)}'^2)
\end{gathered}
\end{equation}

\begin{figure}[hpt!]
    \centering
    \begin{subfigure}[b]{0.4\textwidth}
        \includegraphics[width=\textwidth]{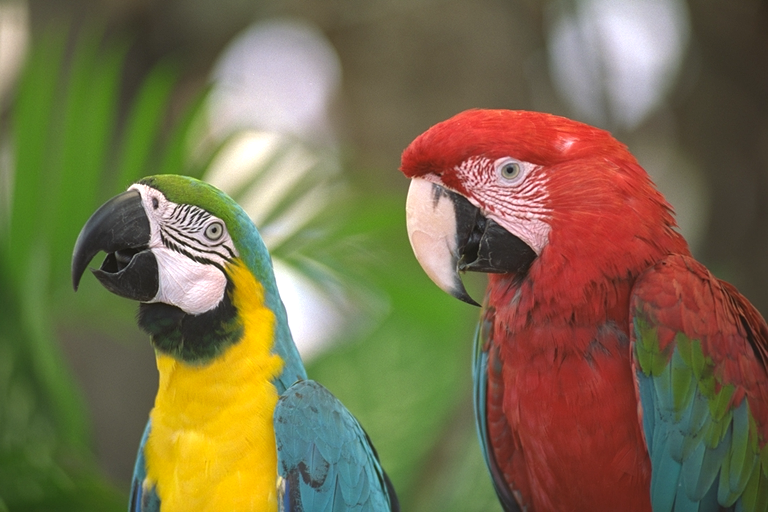}
        \caption{Natural image `parrots'}
        \label{fig:res(a)}
    \end{subfigure} 
    ~ 
    \begin{subfigure}[b]{0.4\textwidth}
        \includegraphics[width=\textwidth]{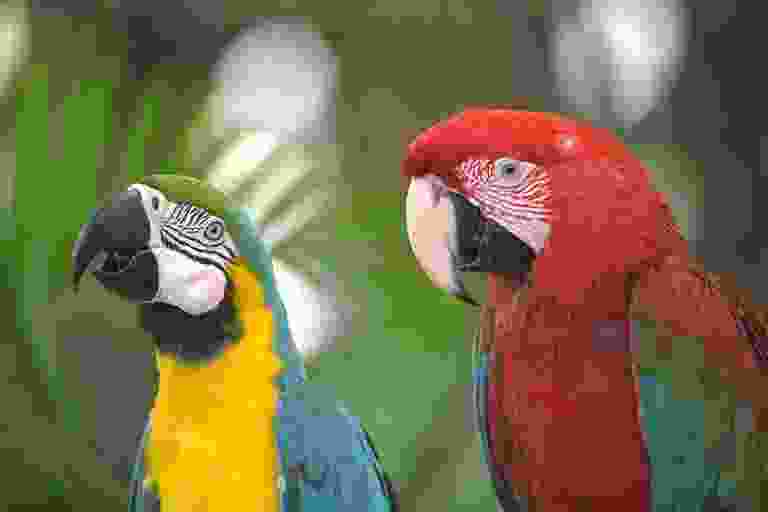}
        \caption{JPEG-compressed `parrots'}
        \label{fig:res(b)}
    \end{subfigure}  \\
    \begin{subfigure}[b]{0.8\textwidth}
        \includegraphics[width=\textwidth]{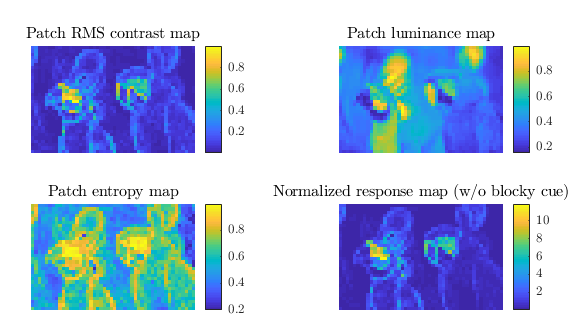}
        \caption{RMS contrast, luminance, entropy response for image (a)}
        \label{fig:res(c)}
    \end{subfigure} 
    \begin{subfigure}[b]{0.8\textwidth}
        \includegraphics[width=\textwidth]{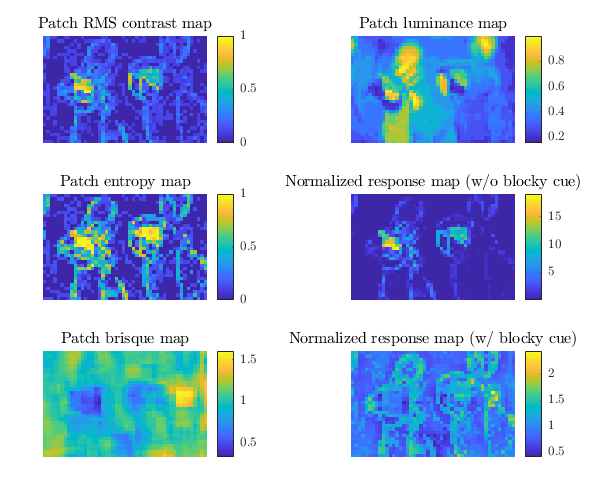}
        \caption{RMS contrast, luminance, entropy, blockiness response for image (b)}
        \label{fig:res(d)}
    \end{subfigure}
    \caption{Proposed image patch response features}
    \label{fig:res}
\end{figure}

Where, $\mathcal{W}_{ik(t)}$ is the response variable with expectation $\mathcal{E}(C_i,L_i,H_i)$ and $\mathcal{N}_{ik(t)}$ is the simulated response noise. $C_i$, $L_i$ and $H_i$ are the saliency-related features defined as RMS contrast, average luminance and image entropy calculated for each patch $P_i$. $\alpha$ is a constant to scale the response map for the whole image to have mean one, whereas $\beta,\gamma,\eta$ are weighting constants which can be fit to data. $N_{ik(t)}$ is sample of internal noise drawn from a normal distribution with standard deviation proportional to $1/d_{ik(t)}'$ and we ignore external noise for simplicity. Fig. \ref{fig:res(a)} shows an example image `parrots.bmp' from LIVE image quality database \cite{sheikh2006statistical} and Fig. \ref{fig:res(c)} displays the defined three types of response maps as well as the overall normalized response map. It can be seen that this response map is highly correlated to visual saliency.

In many images or video communication applications, images are often needed to be compressed to save streaming bandwidth. However, a large compression of the image will definitely result in a huge drop in visual quality. Blocking or ringing artifacts are the most common visual decay coming from modern image coding standards such as JPEG and JPEG2000. Thus, given a compressed image with noticeable blocking or ringing artifacts, like Fig. \ref{fig:res(b)}, it is natural that we should consider both the original response map, defined in Eq. \ref{eq2}, as well as the blocking artifact attractor. However, since there didn't exist any good metric to evaluate the perceived blocking or ringing artifacts for a given image patch, we might as well use the local BRISQUE \cite{mittal2012no} value to approximate the local visual response of blockiness or ringingness. It is somewhat reasonable since BRISQUE metric can considerably capture the image quality degradation, including compression artifact using Natural Scene Statistics features as well as a regression approach. We prefer BRISQUE than NIQE \cite{mittal2013making} since BRISQUE is trained and predicted by SVR \cite{smola2004tutorial} under the supervision of human score, therefore carrying more stabilization and linearity than NIQE (In other words, NIQE is highly nonlinear and unstable). From the `Patch brisque map' shown in Fig. \ref{fig:res(d)}, we can see that the `Patch brisque map' is fairly correlated with the annoyance of blocking/ringing artifacts for local image patches. Therefore, considering all the four factors that drive our attention simultaneously and jointly, the response variable $\mathcal{W}_{ik(t)}$ in Eq. \ref{eq2} can be further modulated by the `Blockiness' index $B_i$ modeled by local BRISQUE value. This gives us the following Eq. \ref{eq3}. 

\begin{equation}
\begin{gathered}
\label{eq3}
\mathcal{W}_{ik(t)}=\mathcal{E}(C_i,L_i,H_i,B_i)+\mathcal{N}_{ik(t)},\ \forall i=1,...,M \\
\mathcal{E}(C_i,L_i,H_i,B_i)=\alpha\cdot \mathcal{N}ormalize( C_i^\beta\cdot L_i^\gamma\cdot H_i^\eta)\cdot B_i^\tau \\
\mathcal{N}_{ik(t)}\sim \mathcal{N}(0,\ 1/d_{ik(t)}'^2)
\end{gathered}
\end{equation}
 
\subsection{Bayesian Updating Method}
\label{ssec:bayes}

Determining the next fixation needs the requirements for the ideal Bayesian searcher: optimal updating posterior distribution, and optimal selection of successive fixation locations. These two parts are the most critical components in the flowchart Fig. \ref{fig:2}. First, we discuss how to optimally update the posterior probability density based on current fixation $T$. 

Suppose $\mathcal{H}_{i}$ is the hypothesis that the potential interesting target position is at the $i$th location (Here $i\in \mathcal{M}$). By Bayes' formula, the posterior $p_{H_i}$ (use $p_i$ for short) is proportional to the multiplication of the prior probability at location $i$ and the joint evidence likelihood of all the block responses at all possible $T$ fixation locations. Let the vector $\boldsymbol{\mathcal{W}(t)}$ represent all the block responses collected at fixation $t$: $(\mathcal{W}_{1k(t)},...,\mathcal{W}_{Mk(t)})$. Given current fixation $T$, we can update the posterior as: 

\begin{equation}
\label{eq4}
p_i(T)\propto prior(i)\cdot p_{i}(\boldsymbol{\mathcal{W}(1)},...,\boldsymbol{\mathcal{W}(T)}|i)
\end{equation}

Where, if independent across fixations, the posterior of $\mathcal{H}_{i}$ is updated as,

\begin{equation}
\label{eq5}
p_{i}(T)\propto prior(i)\cdot \prod^T_{t=1}\prod^M_{i=1}p(\mathcal{W}_{ik(t)}|i) 
\end{equation}

Najemnik and Geisler have shown in the supplement of their paper \cite{najemnik2005optimal} that the Eq. \ref{eq5} is equivalent to updating by multiplying the running summation of the weighted response at $i$th location from all possible fixations $t=1,...,T$, i.e.,

\begin{equation}
\label{eq6}
p_{i}(T)\propto prior(i)\cdot \exp{\big(\sum_{t=1}^T d'^2_{ik(t)}\mathcal{W}_{ik(t)}\big)}
\end{equation}

Where $t$ is the fixation number, and $d'^2_{ik(t)}$, $\mathcal{W}_{ik(t)}$ are the retinal visibility and response variable at patch location $i$ when the fixation is at display location $k(t)$. $d'^2_{ik(t)}$ and $\mathcal{W}_{ik(t)}$ have already been explicitly defined in Section \ref{ssec:vis} and \ref{ssec:patch}. The second term in Eq. \ref{eq6} can be seen as the joint likelihood of the patch response from all potential fixations $k(t)$. Note that Eq. \ref{eq6} is for the case in which both hypotheses are mutually exclusive and independent with both stimulus noise and internal noise statistically independent over time \cite{najemnik2005optimal}.

Here the priors for each fixation is initialized to be equal (i.e., $prior(i)=1/M,\ i=1,...,M$). Considering the `inhibition of return' \cite{klein2000inhibition}, however, we thereby suppress the priors in the neighborhood of the most recent historical fixations (suppose the time interval for each eye saccade is 250 ms). For consistency, we use the same function and parameters as we used for the visibility map in Eq. \ref{eq1}. Hence, the priors are modulated by `inhibition of return' effects as shown in Eq. \ref{eq7}, where the number $n$ is set as the inhibition number of most recent history fixations. According to \cite{pylyshyn1989role}, we use a linearly decreasing weighting $\alpha$ proportional to how recent the historical fixation is to weight the `inhibition effects', as shown in Eq. \ref{eq8}. Here we set the inhibition history as $n=8$, i.e., assuming the searchers can remember the last $2$ seconds in the free-looking situation, and the `inhibition effects' descends linearly as weighted by $\{1/8,2/8,...1\}$.

\begin{equation}
\label{eq7}
prior_{inhib}(i):=prior(i)\cdot\prod^T_{t=T-n}\Big\{1-\boldsymbol{\alpha}\cdot \exp{\Big[-\frac{\epsilon(i,k(t))^2}{2\sigma^2}\Big]}\Big\}
\end{equation}

\begin{equation}
\label{eq8}
\boldsymbol{\alpha}=1-\frac{T-t}{n}\in \Big\{\frac{1}{n},\frac{2}{n},...,1\Big\}
\end{equation}

Thus, the posterior for each potential target location is optimally updated by multiplying the history-suppressed priors and the joint likelihood of the current target patch from each potential fixation. That is, the overall posterior is updated according to Eq. \ref{eq9}. Note that Eq. \ref{eq9} should be normalized by the summation of posteriors among all potential target locations, i.e., $\sum^M_{j=1}p_{j}(T)$, in order to make it a probability measure.

\begin{equation}
\label{eq9}
p_{i}(T)\propto prior_{inhib}(i)\cdot \exp{\big(\sum_{t=1}^T d'^2_{ik(t)}\mathcal{W}_{ik(t)}\big)}
\end{equation}

\subsection{Optimal Selection of Next Fixation}
\label{ssec:select}

Since here we are not conducting a target search task, we don't set any probability criterion for stopping the searching chain; instead, we always use the searcher to iteratively compute the next optimal fixation selection unless the iterations break off. We select three representative arts of Bayesian ideal searcher, dubbed Maximum-a-posteriori (MAP for short) \cite{findlay1982global}, \cite{eckstein2001quantifying}, Entropy Limit Minimization (ELM for short) \cite{najemnik2009simple}, and normalized-ELM search (nELM for short) \cite{abrams2015visual} under the ideal Bayesian probabilistic framework (Fig. \ref{fig:2}) for simulations. Now, we present the optimizing strategies as well as their corresponding reward functions for each of them as follows.  

To compute the optimal next fixation point, $k_{opt}{(T+1)}$, the MAP searcher always fixates the location with maximum posterior probability after the optimal Bayesian updating process, i.e., 

\begin{equation}
\label{eq10}
k_{opt}^{(MAP)}(T+1)=\argmax_i\Big\{p_{i}(T)\Big\}
\end{equation}

Where the $p_i(T)$ is the posterior probability distribution updated by Eq. \ref{eq9}. According to the equation, the MAP searcher always fixates the scene location containing the most salient objects since these areas would produce more significant responses. 

The ideal searcher \cite{najemnik2005optimal}, however, considers each possible next fixation and picks the location that, given its knowledge of current posterior probabilities and visibility map, will maximize the probability of correctly identifying the location of the target after the fixation, shown in Eq. \ref{eq11}.

\begin{equation}
\label{eq11}
k_{opt}^{(Ideal)}(T+1)=\argmax_{k(T+1)}\Big\{\sum_{i=1}^Np_{i}(T)p(C|i,k(T+1))\Big\}
\end{equation}

\begin{algorithm}[!t]
\footnotesize
\caption{The procedures of each simulated search trial among three searchers \{MAP, ELM, nELM\} respectively}
\label{algo}
\begin{algorithmic}[1]
\State Fixation begins at the center of the image (Center prior).
\State Initialize visibility map $d_{ik(t)}'$ by Eq. \ref{eq1} at each possible fixation location $k(t)$.
\State \textbf{Parallel encoding of the image}: At each fixation $k(t)$, an independent Gaussian noise sample $\mathcal{N}_{ik(t)}$ (with zero mean and variance $1/d_{ik(t)}'^2$) is generated for each target patch $P_i$. Thus, the response variable $\mathcal{W}_{ik(t)}$ at each patch $P_i$ from current fixation $t$ is computed by Eq. \ref{eq2} or \ref{eq3}. (for pristine image like Fig. \ref{fig:res(a)}, we use Eq. \ref{eq2}; for compressed image like Fig. \ref{fig:res(b)}, we use Eq. \ref{eq3}).
\State \textbf{Optimal Bayesian updating}: after fixation $T$ is chosen, the posterior probability at each image patch location $i$ is updated by the Bayes formula Eq. \ref{eq9}, where the priors are initialized by Eq. \ref{eq7} as follows. In other words, optimal integration of information across fixations is achieved by keeping a running summation of weighted response at each potential location. 
$$
p_{i}(T)\propto prior_{inhib}(i)\cdot \exp{\big(\sum_{t=1}^T d'^2_{ik(t)}\mathcal{W}_{ik(t)}\big)}, 
$$
$$
prior_{inhib}(i):=prior(i)\cdot\prod^T_{t=T-n}\Big\{1-\boldsymbol{\alpha}\cdot \exp{\Big[-\frac{\epsilon(i,k(t))^2}{2\sigma^2}\Big]}\Big\}
$$
\State \textbf{Optimally choose the next fixation}: to compute the optimal next fixation point, $k_{opt}(T+1)$, the searcher considers each possible next fixation and picks the location with the highest expected value, a product of sensory evidence and potentially earned rewards, which are defined differently: 

\textbf{SWITCH} `method'$\in$ \{`MAP',`ELM',`nELM'\} \textbf{DO}

\textbf{CASE} `MAP':

\[k_{MAP}(T+1)=\argmax_i\Big\{p_{i}(T)\Big\}\]

\textbf{CASE} `ELM':

\[k_{ELM}(T+1)=\argmax_{k(T+1)}\Big\{\sum_{i=1}^Mp_{i}(T)d_{ik(T+1)}^2\Big\}\]
        
\textbf{CASE} `nELM':

\[p_{i}^{(Norm)}(T):=p_{i}(T)/C_i,\ for\ i=1,...,M \]
\[k_{nELM}(T+1)=\argmax_{k(T+1)}\Big\{\sum_{i=1}^Mp_{i}^{(Norm)}(T)d_{ik(T+1)}^2\Big\}\]
\State The process jumps back to \textbf{Step 3} and repeats the search trial \textbf{Step 3-5} to predict the next optimal fixation location.

\end{algorithmic}
\end{algorithm}

Where the ideal searcher Eq. \ref{eq11} tries to maximize the accuracy in the target-searching task. However, given the computational complexity in ideal searcher \ref{eq11}, Najemnik and Geisler then derived a simple heuristic model called entropy limit minimization (ELM) searcher \cite{najemnik2009simple} which produces near-optimal fixation selection and produces fixation statistics similar to human. They proved that, under some simple summation rule assumptions, the expected information gain of fixation $k(T+1)$ (Eq. \ref{eq12(1)}) with regard to previous fixation $k(T)$ is equivalent to linearly filtering the current posterior across the possible target locations (Eq. \ref{eq12(4)}). They derived the Eq. \ref{eq12(4)} using Eq. \ref{eq12(1)},\ref{eq12(2)},\ref{eq12(3)}. Therefore, the ELM searcher chooses the optimal next fixated location $k(T+1)$ as the one that maximized the expected information gain as approximated by the summation formula shown in Eq. \ref{eq13}, where the posteriors are updated in the same scheme Eq. \ref{eq9} as in MAP.

\begin{equation}
\label{eq12(1)}
\mathbf{E}\big[\Delta H(T+1)|k(T+1)\big]=\mathbf{E}\big[H(T+1)|k(T+1)\big]-H(T) 
\end{equation}

\begin{equation}
\label{eq12(2)}
\mathbf{E}\big[H(T+1)|k(T+1)\big]=-\mathbf{E}\Big[\sum_{i=1}^n P_i(T+1)\log{(P_i(T+1))}|k(T+1)\Big]  
\end{equation}

\begin{equation}
\label{eq12(3)}
P_i(T+1)=\frac{p_i(T)\exp{\big[d^2_{ik(T+1)}\mathcal{W}_{ik(T+1)}}\big]}{\sum_{j=1}^np_j(T)\exp{\big[d^2_{jk(T+1)}\mathcal{W}_{jk(T+1)}}\big]}
\end{equation}

\begin{equation}
\label{eq12(4)}
\mathbf{E}\big[\Delta H(T+1)|k(T+1)\big]\approx\frac{1}{2}\sum_{i=1}^N p_i(T)d^2_{ik(T+1)} 
\end{equation}

\begin{equation}
\label{eq13}
k_{opt}^{(ELM)}(T+1)=\argmax_{k(T+1)}\Big\{\sum_{i=1}^Mp_{i}(T)d_{ik(T+1)}^2\Big\}
\end{equation}

Normalized ELM (nELM) searcher \cite{abrams2015visual}, which tries to take into account the variation of detectability map modulated by local contrast of gaze point, was built on ELM to further generalize for non-stationary natural background. nELM searcher first normalizes the posteriors by local contrast, then picks the fixate location with maximized expected information gain, expressed in Eq. \ref{eq14}.

\begin{equation}
\label{eq14}
\begin{gathered}
p_{i}^{(Norm)}(T):=p_{i}(T)/C_i,\ for\ i=1,...,M \\
k_{opt}^{(nELM)}(T+1)=\argmax_{k(T+1)}\Big\{\sum_{i=1}^Mp_{i}^{(Norm)}(T)d_{ik(T+1)}^2\Big\}
\end{gathered}
\end{equation}

\subsection{MAP, ELM and nELM Searchers Under Distortions}
\label{ssec:sim}

Based on the above assumptions and definitions, the steps in \textbf{Algorithm} \ref{algo} show in detail how the searcher works in each trial recursively. 

\section{Eye Fixation Search Experiments and Results}

\subsection{Test on Stationary Backgrounds}

\begin{figure}[t!]
    \centering
    \begin{subfigure}[b]{0.45\textwidth}
        \includegraphics[width=\textwidth]{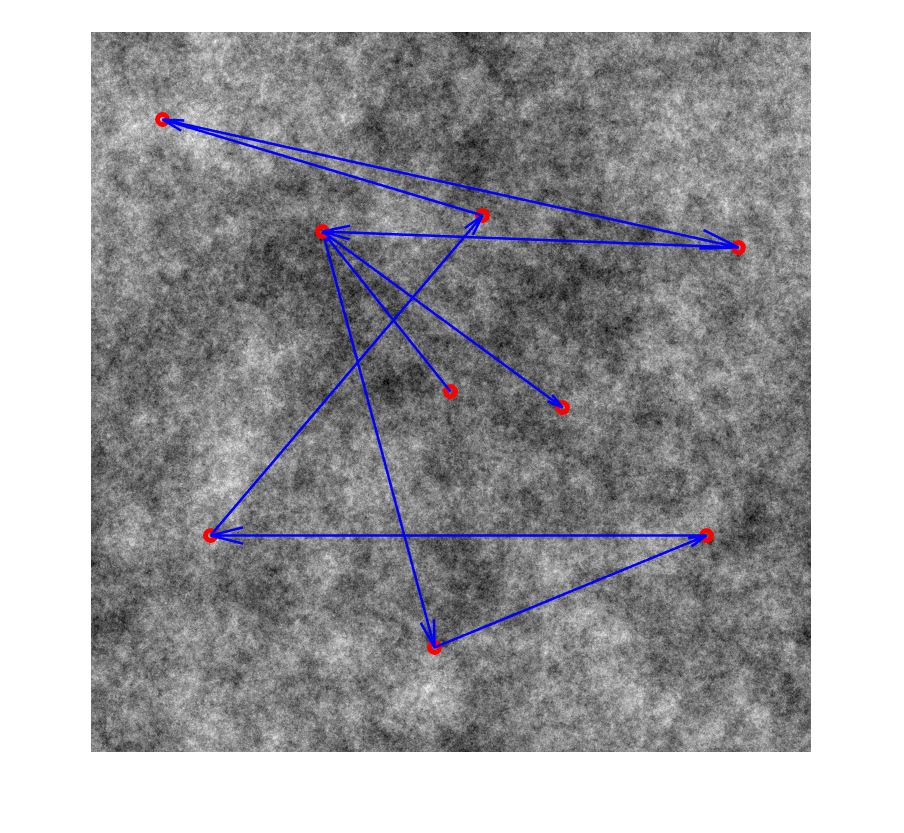}
        \caption{First 10 fixations predicted by MAP}
        \label{fig:5(a)}
    \end{subfigure} 
    \begin{subfigure}[b]{0.45\textwidth}
        \includegraphics[width=\textwidth]{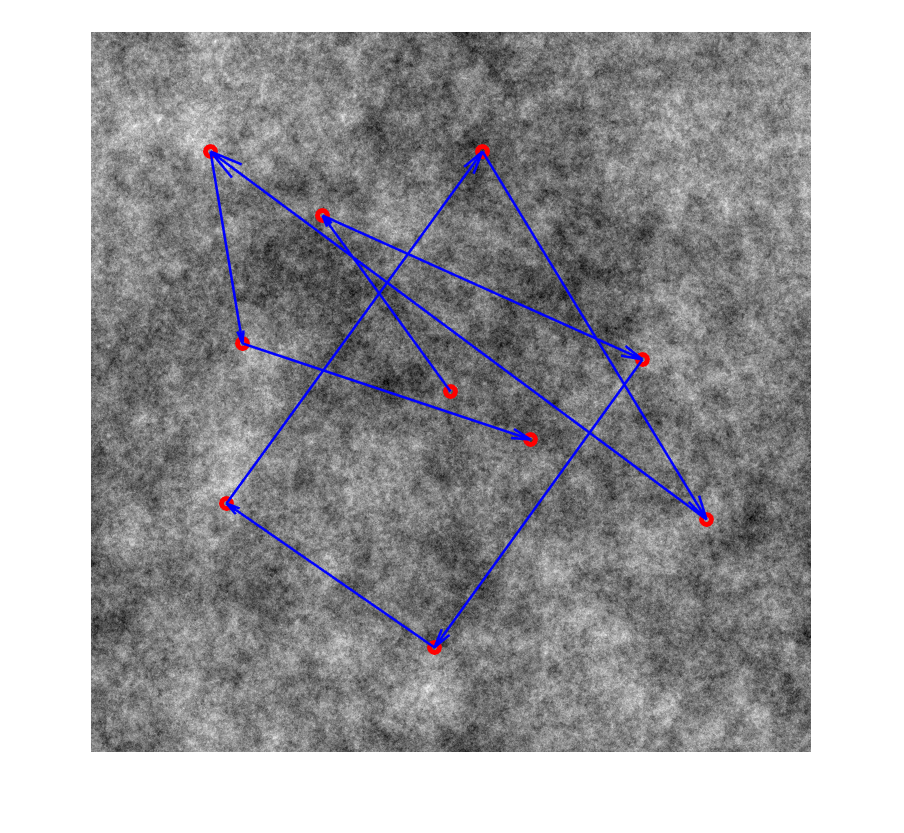}
        \caption{First 10 fixations predicted by ELM}
        \label{fig:5(b)}
    \end{subfigure}  \\
        \begin{subfigure}[b]{0.45\textwidth}
        \includegraphics[width=\textwidth]{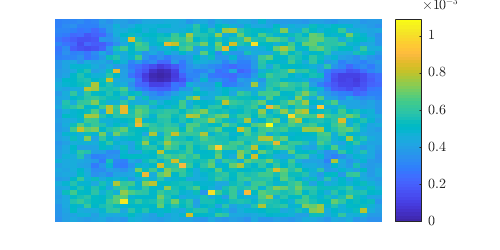}
        \caption{The 10th posterior modulated by `Inhibition of return
        }
        \label{fig:5(c)}
    \end{subfigure}
        \begin{subfigure}[b]{0.45\textwidth}
        \includegraphics[width=\textwidth]{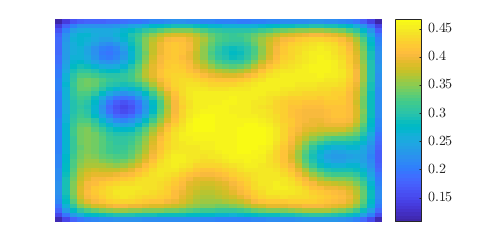}
        \caption{The 10th information gain map modulated by `Inhibition of return'}
        \label{fig:5(d)}
    \end{subfigure}
    
    \caption{Fixation prediction test on $1/f$ noise image}
    \label{fig:5}
\end{figure}

We generate the spatial $1/f$ noise image which has the same spatial power spectra as natural images as the initial test to validate our simulations of the searchers. Note that the $1/f$ image has stationary local contrast, so there is no need to test the nELM algorithm on it. We can see from Fig. \ref{fig:5(a)} and \ref{fig:5(b)} that oftentimes, ELM chooses the same fixation point to maximize the posterior just as what MAP does because information gain map is just the visibility map-blurred posteriors. However, MAP can take a saccade to somewhere near the edge whereas ELM tends only to jump a moderate distance since posteriors nearby are pushed down by `inhibition of return' and information gain at distant locations tends not to be high at all. Fig. \ref{fig:5(d)} shows the history-inhibited information gain map, which has much lower value near the edges as well as the visited locations. 

\subsection{Test on Natural Images}

\begin{figure}[t!]
    \centering
    \begin{subfigure}[b]{0.45\textwidth}
        \centering
        \includegraphics[width=\textwidth]{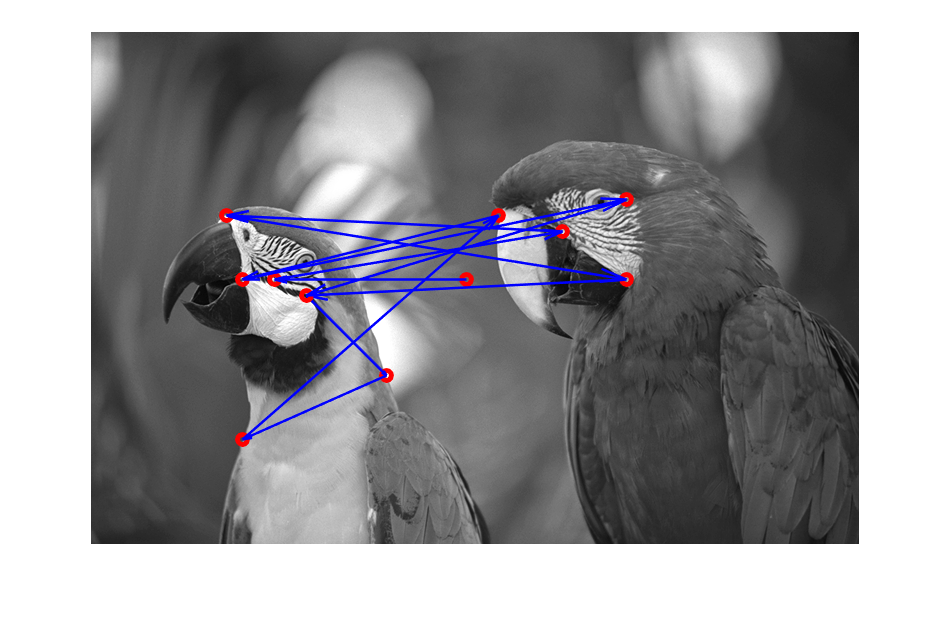}
        \caption{First 12 fixations predicted by MAP on natural image `parrots'}
        \label{fig:6(a)}
    \end{subfigure} 
    \begin{subfigure}[b]{0.45\textwidth}
    \centering
        \includegraphics[width=0.8\textwidth]{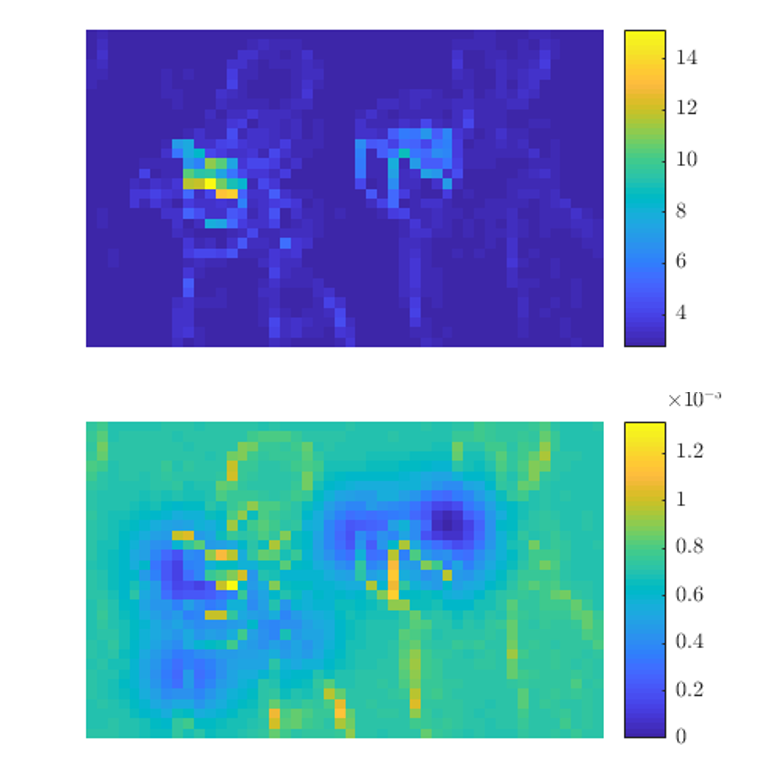}
        \caption{Likelihood map (above) and posteriors (below)}
        \label{fig:6(b)}
    \end{subfigure}  \\
    \begin{subfigure}[b]{0.45\textwidth}
        \centering
        \includegraphics[width=\textwidth]{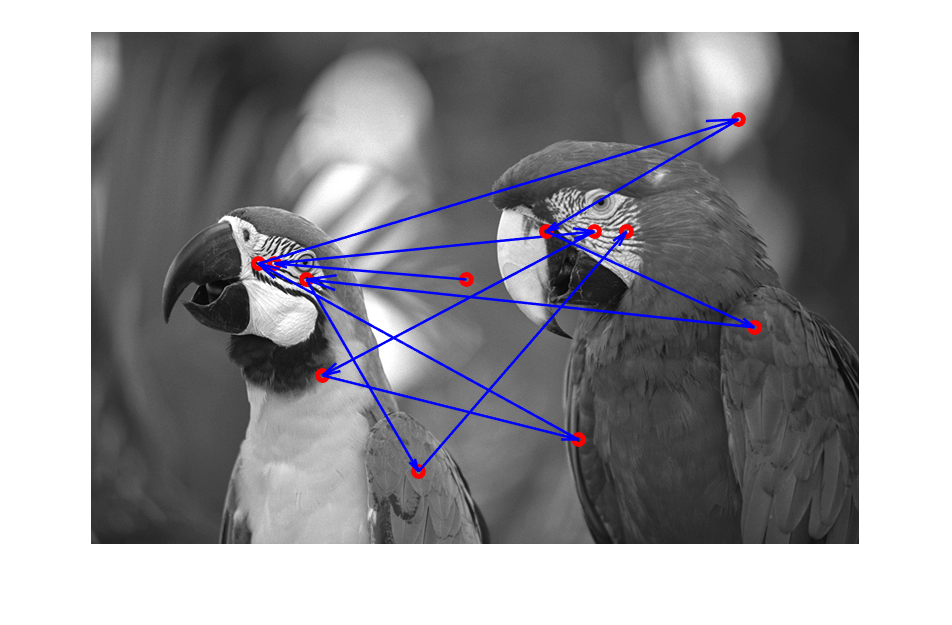}
        \caption{First 12 fixations predicted by ELM on natural image `parrots'}
        \label{fig:6(c)}
    \end{subfigure}
    \begin{subfigure}[b]{0.45\textwidth}
        \centering
        \includegraphics[width=0.8\textwidth]{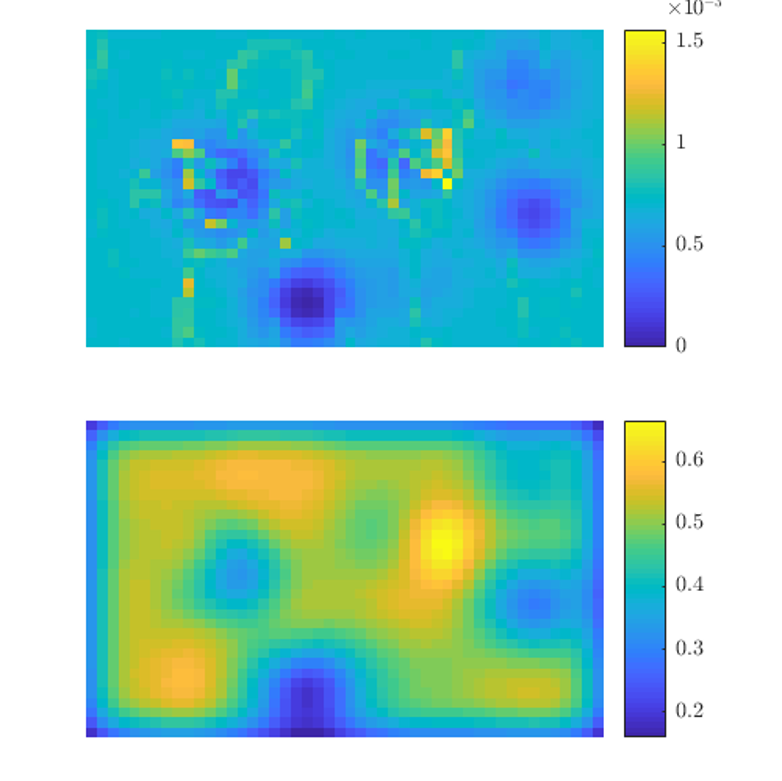}
        \caption{Posteriors (above) and information gain map (below)}
        \label{fig:6(d)}
    \end{subfigure}
    \begin{subfigure}[b]{0.45\textwidth}
        \centering
        \includegraphics[width=\textwidth]{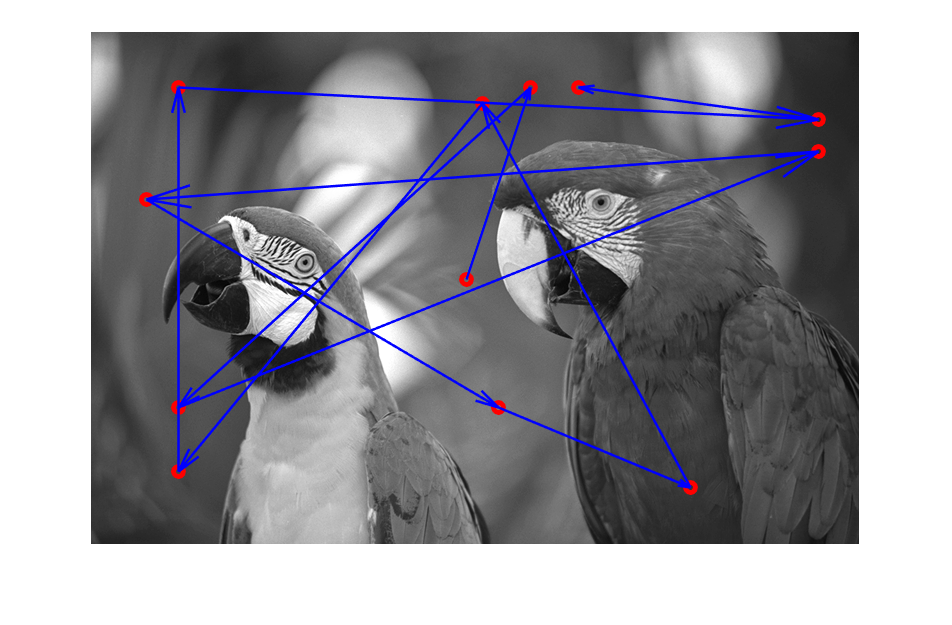}
        \caption{First 12 fixations predicted by nELM on natural image `parrots'}
        \label{fig:6(e)}
    \end{subfigure}
    \begin{subfigure}[b]{0.45\textwidth}
        \centering
        \includegraphics[width=0.8\textwidth]{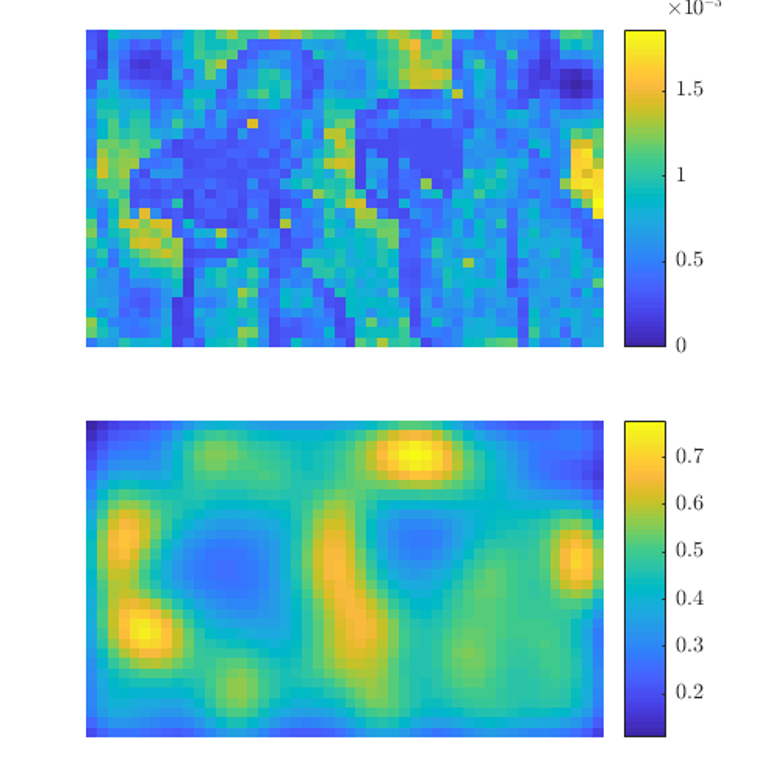}
        \caption{Normalized posteriors (above) and information gain map (below)}
        \label{fig:6(f)}
    \end{subfigure}
    
    \caption{Fixation prediction results on natural image `parrots'}
    \label{fig:6}
\end{figure}

We selected the image `parrots.bmp' from LIVE Image Quality Database \cite{sheikh2006statistical}, on which we test the \textbf{Algorithm} \ref{algo} with $12$ fixations prediction trials. The number of inhibited fixations are set to be $n=8$ with linearly decreasing weights. Fig. \ref{fig:6} shows the search results using MAP (Fig. \ref{fig:6(a)}), ELM (Fig. \ref{fig:6(c)}), and nELM (Fig. \ref{fig:6(e)}) respectively. Figures on the right column of Fig. \ref{fig:6} shows as the examples of likelihood map, posteriors, information gain map, and normalized posteriors both at the last fixation prediction. As we can see from the figures, MAP tends to choose fixations only on most salient areas, whereas ELM sometimes fixates at locations with lower posterior probability but with higher information gain. Note that in Fig. \ref{fig:6(e)}, we regard the predicted fixations by nELM not good enough since almost all the saccades land in uninteresting points and the saccades distance is relatively long. That is possible because we simply normalize the posteriors by dividing the local contrast, which would result in the posterior of some area with lower contrast becoming unexpectedly larger after normalization. This problem may be improved by fine-tuning the normalization function based on some subjective human studies.

\begin{figure}[t!]
    \centering
    \begin{subfigure}[b]{0.45\textwidth}
        \centering
        \includegraphics[width=\textwidth]{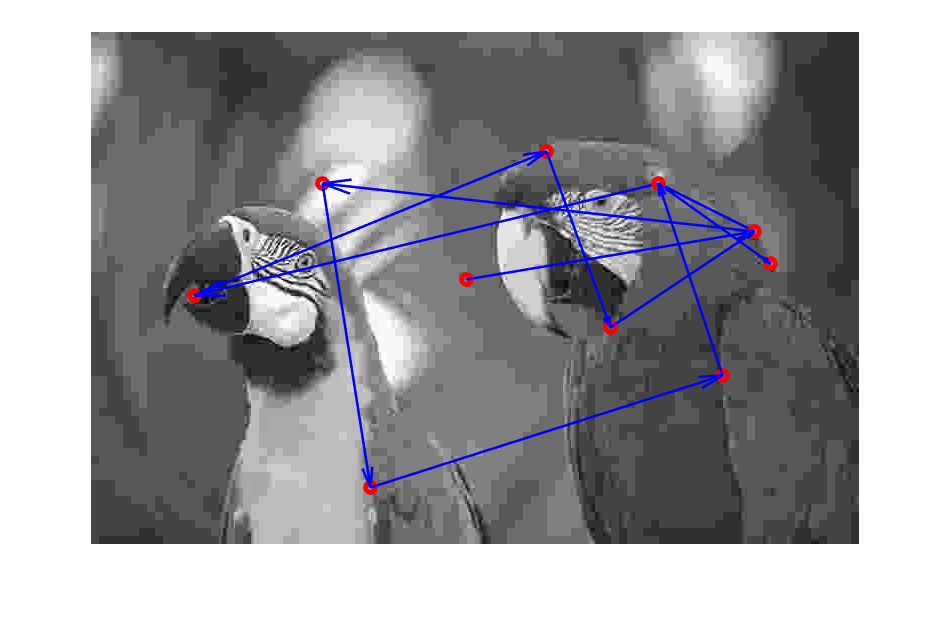}
        \caption{First 12 fixations predicted by MAP on compressed natural image `parrots'}
        \label{fig:7(a)}
    \end{subfigure} 
    \begin{subfigure}[b]{0.45\textwidth}
    \centering
        \includegraphics[width=0.8\textwidth]{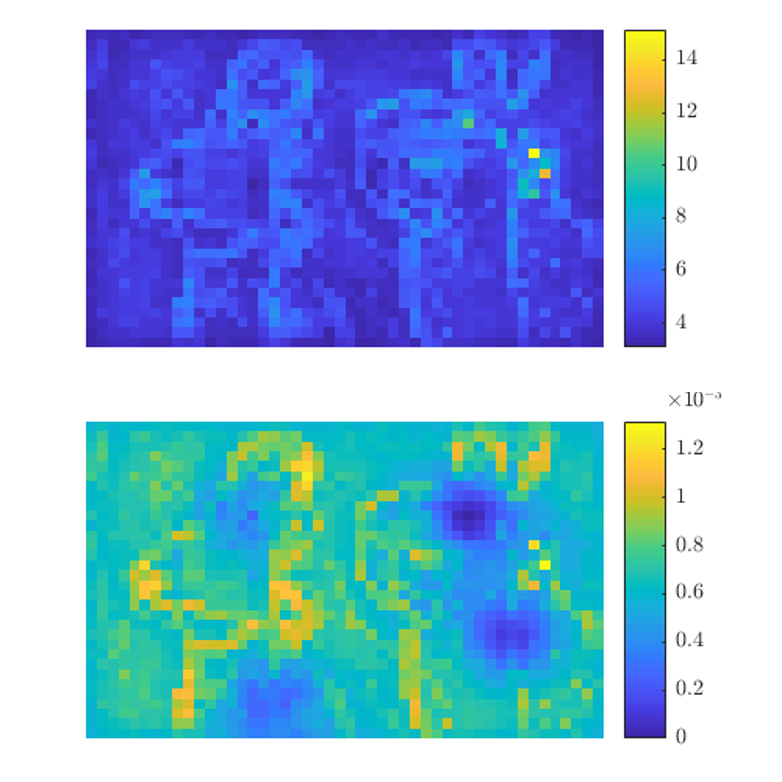}
        \caption{Likelihood map (above) and posteriors (below)}
        \label{fig:7(b)}
    \end{subfigure}  \\
    \begin{subfigure}[b]{0.45\textwidth}
        \centering
        \includegraphics[width=\textwidth]{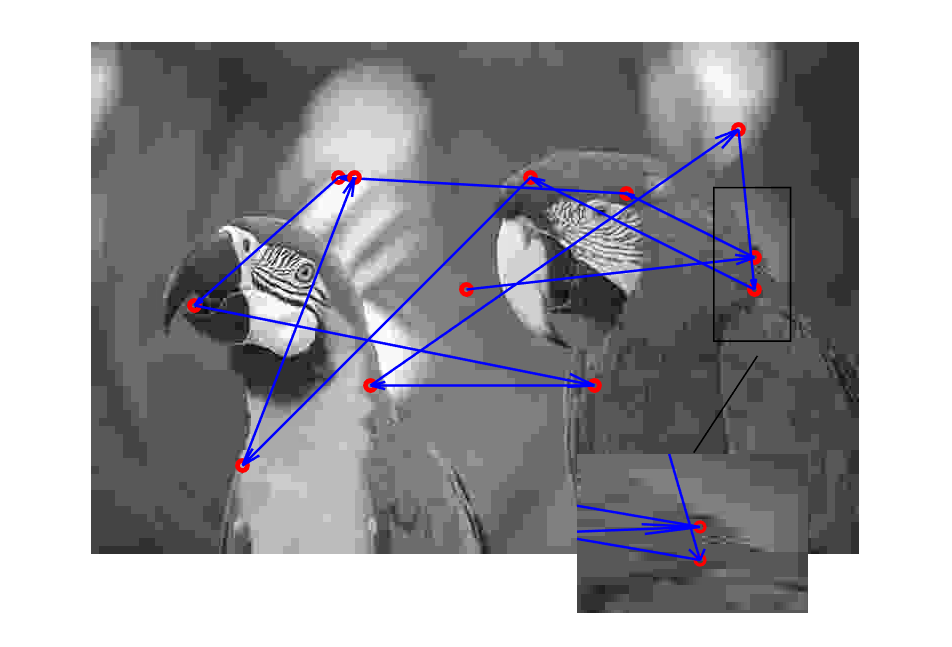}
        \caption{First 12 fixations predicted by ELM on compressed natural image `parrots'}
        \label{fig:7(c)}
    \end{subfigure}
    \begin{subfigure}[b]{0.45\textwidth}
        \centering
        \includegraphics[width=0.8\textwidth]{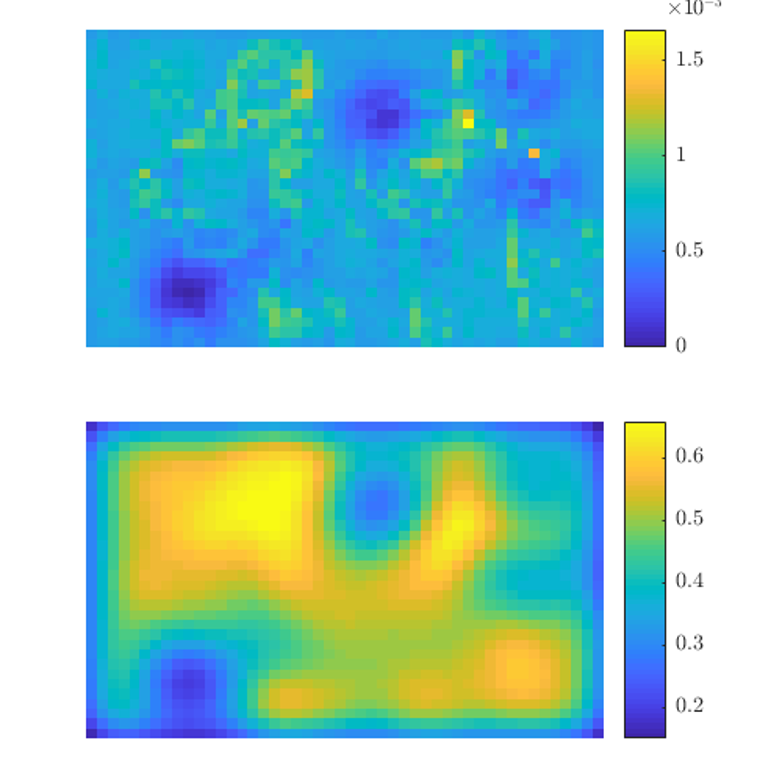}
        \caption{Posteriors (above) and information gain map (below)}
        \label{fig:7(d)}
    \end{subfigure}
    \begin{subfigure}[b]{0.45\textwidth}
        \centering
        \includegraphics[width=\textwidth]{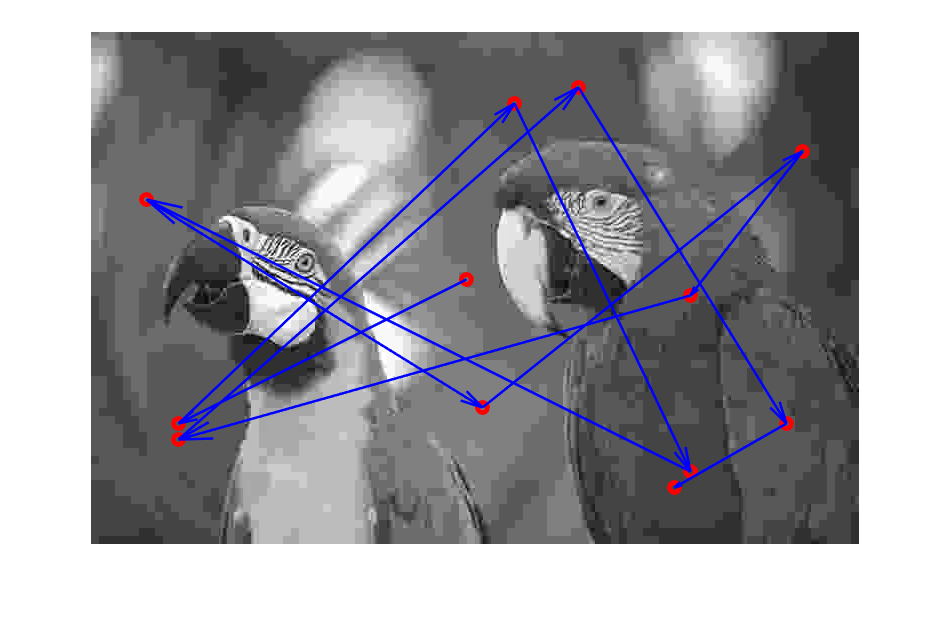}
        \caption{First 12 fixations predicted by nELM on compressed natural image}
        \label{fig:7(e)}
    \end{subfigure}
    \begin{subfigure}[b]{0.45\textwidth}
        \centering
        \includegraphics[width=0.8\textwidth]{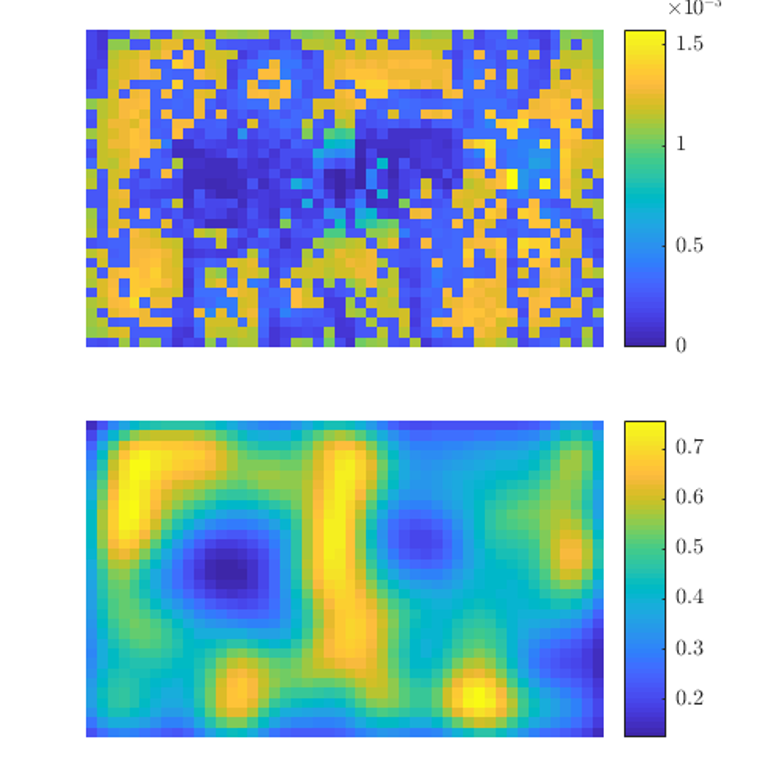}
        \caption{Normalized posteriors (above) and information gain map (below)}
        \label{fig:7(f)}
    \end{subfigure}
    
    \caption{Fixation prediction results on compressed natural image `parrots'}
    \label{fig:7}
\end{figure}

\begin{figure}[t!]
    \centering
    \begin{subfigure}[b]{0.4\textwidth}
        \centering
        \includegraphics[width=\textwidth]{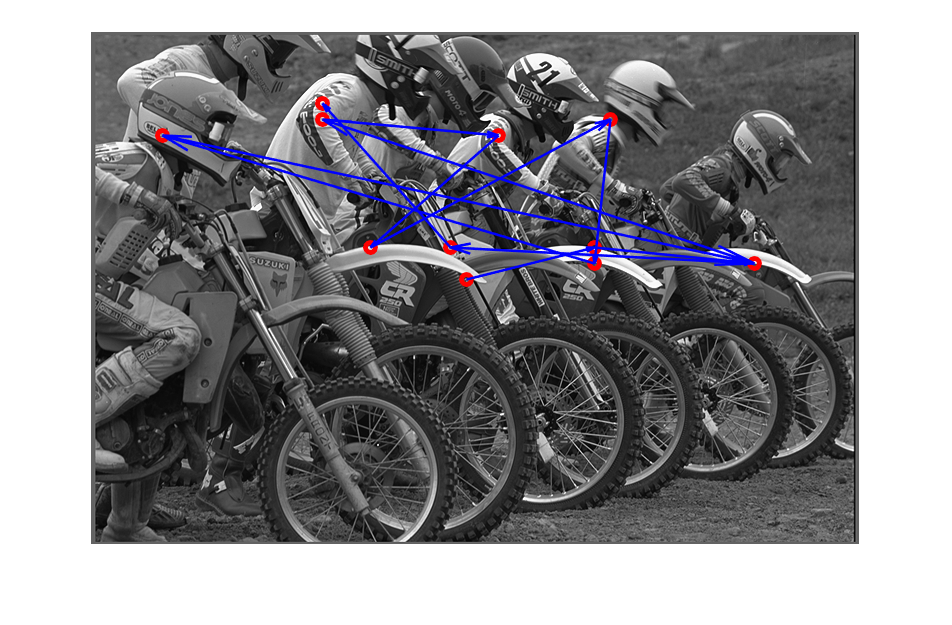}
        \caption{ELM on `bikes.bmp'}
        \label{fig:8(a)}
    \end{subfigure} 
    \begin{subfigure}[b]{0.4\textwidth}
    \centering
        \includegraphics[width=\textwidth]{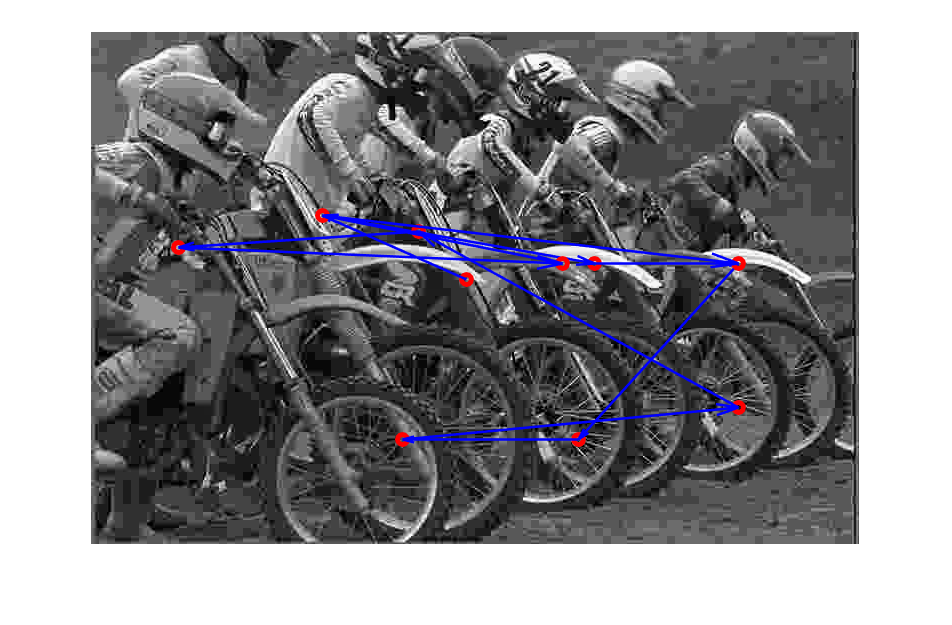}
        \caption{ELM on compressed `bikes.jpg'}
        \label{fig:8(b)}
    \end{subfigure}  \\
    \begin{subfigure}[b]{0.4\textwidth}
        \centering
        \includegraphics[width=\textwidth]{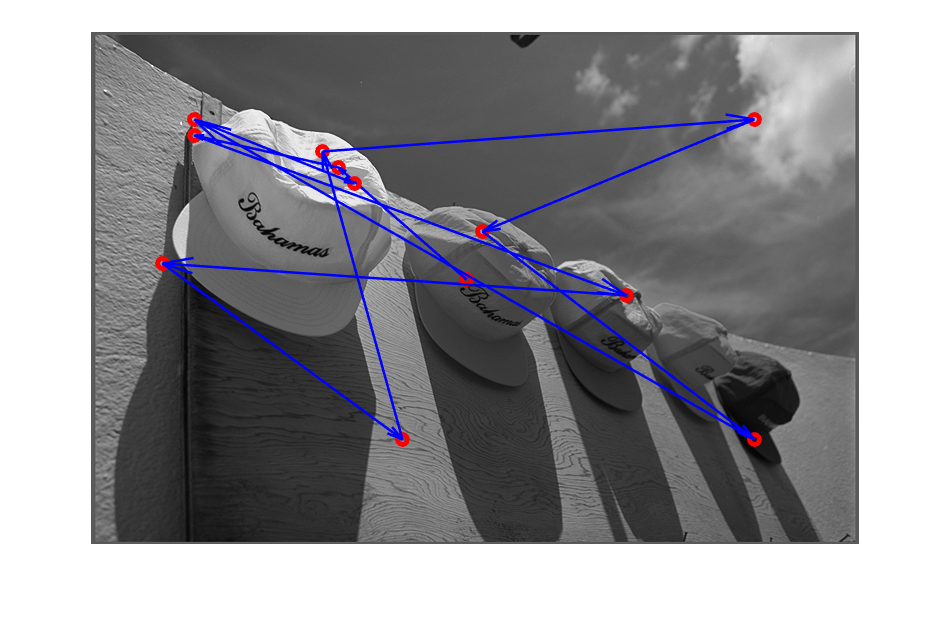}
        \caption{ELM on `caps.bmp'}
        \label{fig:8(c)}
    \end{subfigure}
    \begin{subfigure}[b]{0.4\textwidth}
        \centering
        \includegraphics[width=\textwidth]{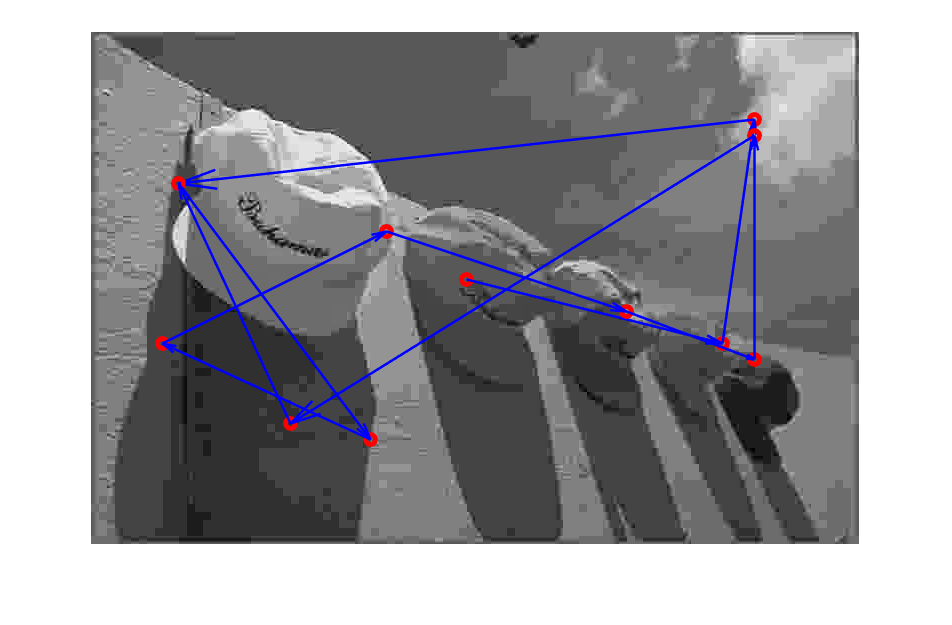}
        \caption{ELM on compressed `caps.jpg'}
        \label{fig:8(d)}
    \end{subfigure}
    \begin{subfigure}[b]{0.4\textwidth}
        \centering
        \includegraphics[width=\textwidth]{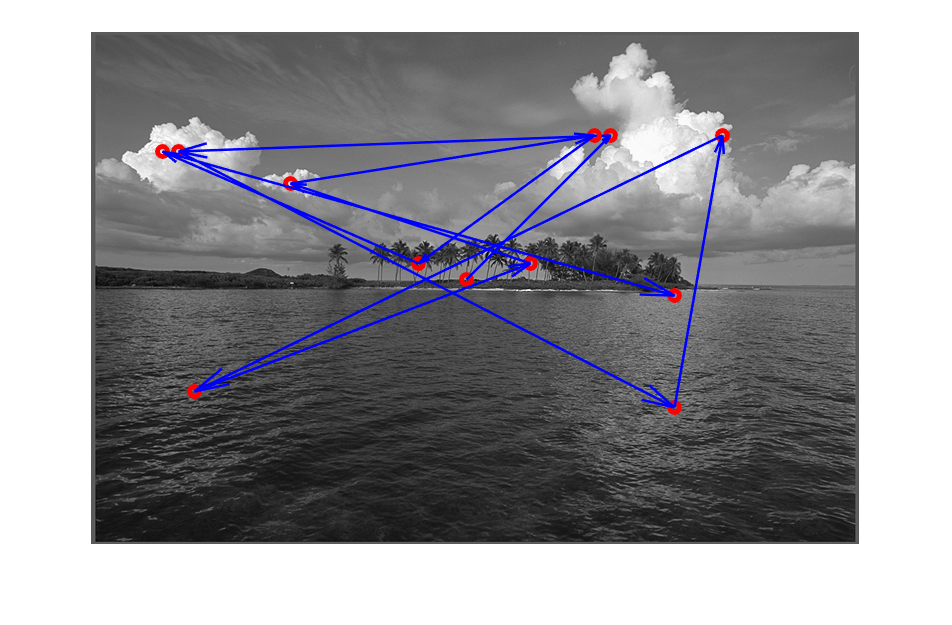}
        \caption{ELM on `ocean.bmp'}
        \label{fig:8(g)}
    \end{subfigure}
    \begin{subfigure}[b]{0.4\textwidth}
        \centering
        \includegraphics[width=\textwidth]{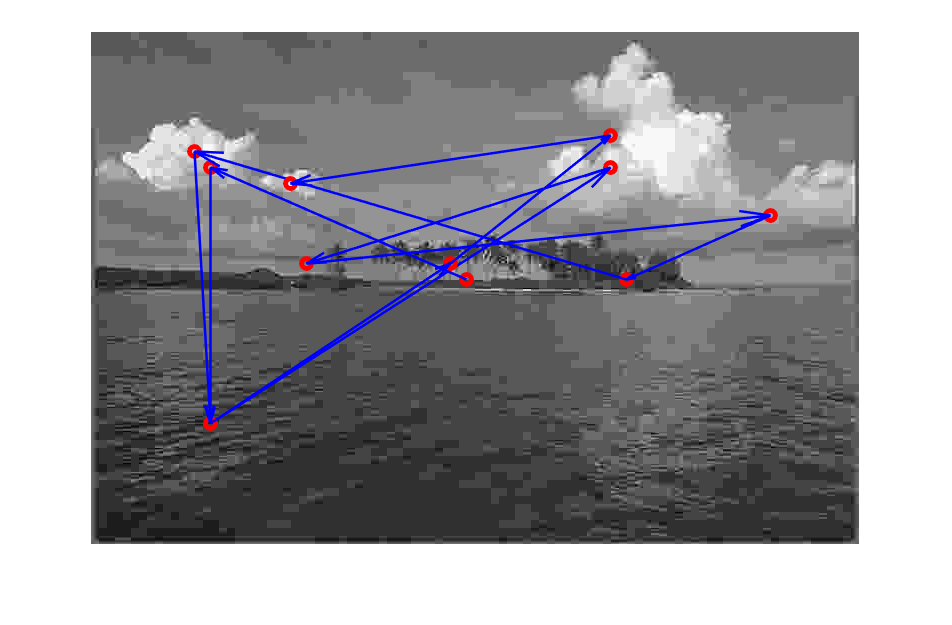}
        \caption{ELM on compressed `ocean.jpg'}
        \label{fig:8(h)}
    \end{subfigure}
    \begin{subfigure}[b]{0.4\textwidth}
        \centering
        \includegraphics[width=\textwidth]{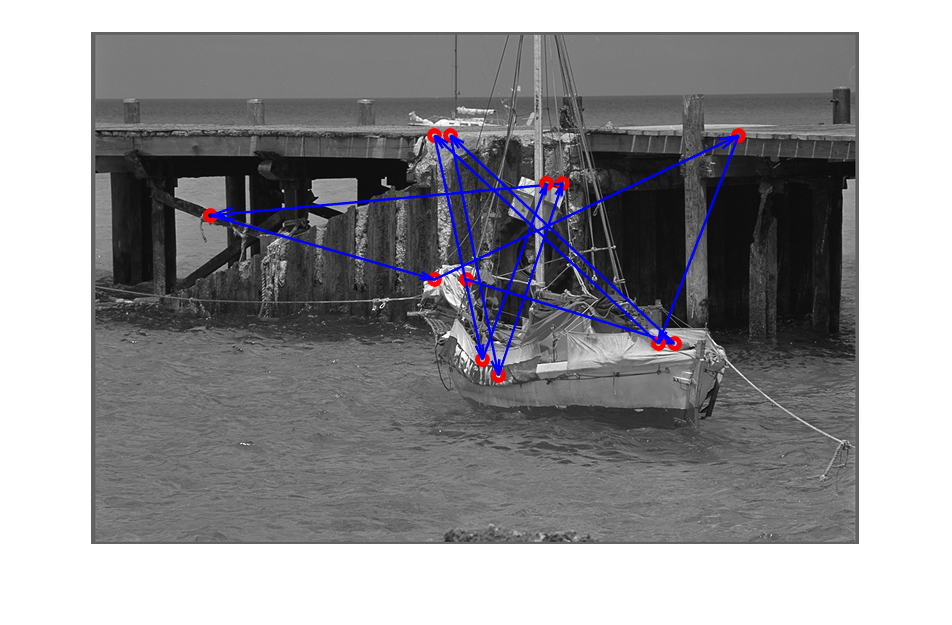}
        \caption{ELM on `sailing4.bmp'}
        \label{fig:8(i)}
    \end{subfigure}
    \begin{subfigure}[b]{0.4\textwidth}
        \centering
        \includegraphics[width=\textwidth]{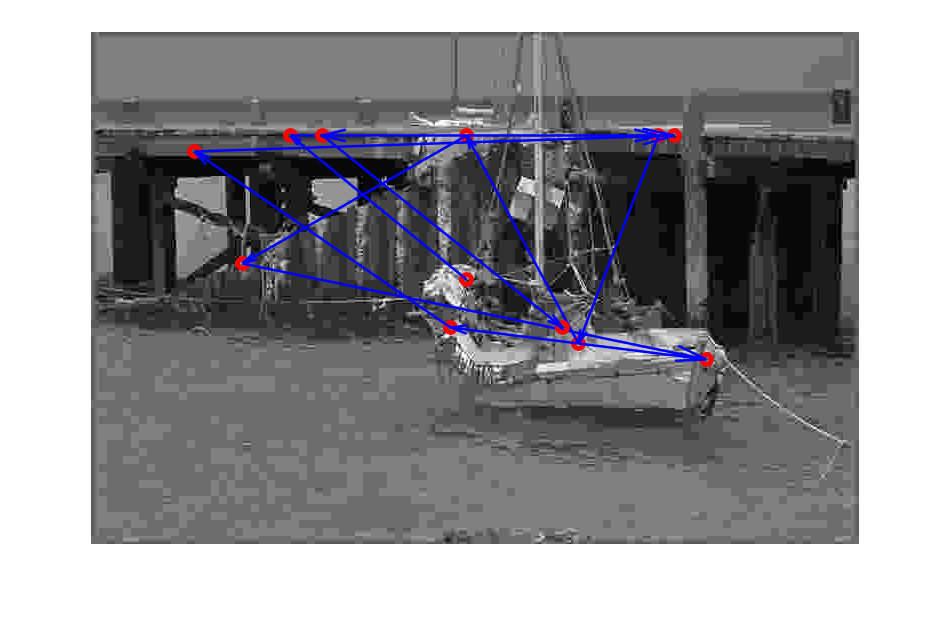}
        \caption{ELM on compressed `sailing4'}
        \label{fig:8(j)}
    \end{subfigure}
    \caption{More eye fixation results using ELM searcher on images from LIVE database}
    \label{fig:8}
\end{figure}

\subsection{Test on Compressed Natural Images}

Since our goal is to study the distracting effects of image compression artifacts on predicted scanpaths, we used the same image `parrots.bmp' and compressed it with $Quality=5$ using $imwrite$ function in MATLAB. After obtaining the patch response map, we did histogram equalization to make the distribution more dispersed and smooth. Then we used Eq. \ref{eq3} with parameters $(\beta,\gamma,\eta,\tau)=(1,1,1,1)$ to calculate the response and accordingly updated the posteriors. Fig. \ref{fig:7} shows the first $12$ fixations predicted by MAP (Fig. \ref{fig:7(a)}), ELM (Fig. \ref{fig:7(c)}), and nELM (Fig. \ref{fig:7(e)}). We observed that both searchers are distracted by the noticeable blocking or ringing artifacts to some extent, while each of the searchers is still identifiable by its own searching fashion. The zoomed image region in Fig. \ref{fig:7(c)}, which is full of notably perceptible blocking artifacts, is selected as the first fixate location by ELM, therefore verifying our expectations that noticeable image quality degradation truly distract our visual attention against salient areas.

\section{Conclusion and Discussion}

Conclusively, this paper explores the Bayesian probabilistic framework shown in Fig. \ref{fig:2} and slightly modifies it to predict human eye fixations in the context of free-looking at natural images suffered from compression artifacts. We defined the response variables of a distortion-free image patch in a similar way with Najemnik and Geisler \cite{najemnik2005optimal}. Given a compressed image with noticeable distortions, we further defined the `blockiness response map' as an approximation of how salient the `blocking artifact' presents to a human observer and then used this distortion map to compete with traditional saliency maps (i.e., distortion-free response map). Integrating this additional visual factor dubbed `blockiness' to the basic response variable, we thereby simulated three classic fixation searchers under a Bayesian probabilistic framework, which are MAP \cite{findlay1982global}, ELM \cite{najemnik2009simple}, and nELM \cite{abrams2015visual} respectively. The experimental results on different images from LIVE Database show that the Bayesian visual search is indeed disturbed by the annoying blocking artifacts. These simulations have presented some evidence for the distracting effects of image distortions on visual search and build a preliminary model based on the Bayesian probabilistic framework for potential subsequent explorations on interactions of visual attention and image quality. 

Lastly, we discuss as follows some deficiencies of the implementations in this paper, as well as their potential improvement: 

\begin{itemize}
    \item \textbf{Accuracy of detectability map.} We assume a simplified model as a Gaussian distribution shown in Fig. \ref{fig:3(a)} for mathematically convenience in which the target detectability is dependent on eccentricity and isotropic. However, this does not seem right since the real retinotopic detectability map looks more like an ellipse and also tuned by background contrast. The implemented searchers in this paper are expected to achieve more human-alike performance if a more accurate retinotopic detectability map is used.
    \item \textbf{Accuracy of redefined response map.} Here we redefine the image response map which indicates the saliency of a patch as the multiplication of patch contrast, luminance, and entropy, which is a very rough approximation; besides, the `blocking response map' is likewise quite loose. One can just improve by leveraging state-of-the-art saliency map algorithm, for example. From my point of view, given a distorted image, how to define the response map is the trickiest part in both searchers within this Bayesian framework since it will directly weight the posteriors, thereby deciding fairly the next optimal fixate location.
    \item \textbf{Effects of `Inhibition of Return'.} We noticed that the `Inhibition of Return' phenomenon exhibits more significant impacts than maximizing the reward function on driving the next eye movement. Specifically, attention is encouraged towards new locations where fixations have not yet visited. Here we simply set the memorized fixation history to be lasting $2$ seconds (i.e., $8$ previous fixations) with descending weights, which needs further consideration.
    \item \textbf{Subjective verification.} Currently, we only show some visual examples of the predicted eye fixations from a Bayesian search framework. However, the accuracy of approximating human fixations has not been justified in this paper. A subjective experiment of human eye fixations on distorted images is expected to be conducted to verify our hypothesis in this paper.
    \item \textbf{Improving video quality models.} Objective video quality models have raised increasingly significant research interest recently \cite{tu2020ugc, chen2020proxiqa}. It would be of great significance to study the effects of eye fixations on the performance of video quality models, especially on the recently popular user-generated content (UGC). We may also adopt a content-adaptive streaming scheme \cite{tu2018content} if human eye fixations can be accurately predicted to save bitrate. 
\end{itemize}

\clearpage

\bibliographystyle{splncs}
\bibliography{egbib}
\end{document}